\documentclass[twocolumn,linenumbers]{aastex631}

\usepackage{amsmath}
\usepackage{comment}

\begin{document}
\nolinenumbers
\title{Signatures of Rapidly Rotating Stars with Chemically Homogeneous Evolution in the First Galaxies  }

\correspondingauthor{Boyuan Liu}
\email{Email: boyuan.liu@uni-heidelberg.de}

\author[0000-0002-4966-7450]{Boyuan Liu}
\affiliation{Institut für Theoretische Astrophysik, Zentrum für Astronomie, Universität Heidelberg, D-69120 Heidelberg, Germany}
\author[0000-0002-7829-776X]{Yves Sibony}
\affiliation{Department of Astronomy, University of Geneva, Chemin Pegasi 51, 1290 Versoix, Switzerland}
\author[0000-0001-6181-1323]{Georges Meynet}
\affiliation{Department of Astronomy, University of Geneva, Chemin Pegasi 51, 1290 Versoix, Switzerland}
\author[0000-0003-0212-2979]{Volker Bromm}
\affiliation{Department of Astronomy, 
University of Texas at Austin, 2515 Speedway, Stop C1400, Austin, TX 78712, USA}
\affiliation{Weinberg Institute for Theoretical Physics, 
University of Texas at Austin, Austin, TX 78712, USA}







\begin{abstract}
\nolinenumbers
The James Webb Space Telescope ({\it JWST}) has revealed an unexpectedly high abundance of UV luminous galaxies at redshifts $z\gtrsim 10$, challenging `standard' galaxy formation models. This study investigates the role of rapidly rotating (massive) stars undergoing chemically homogeneous evolution (CHE) in reconciling this potential tension. These stars are more compact, hotter, and exhibit enhanced UV emission. We find that the rest-frame UV luminosity of star-forming galaxies can be significantly enhanced by a factor of $\sim 3-6$ when CHE stars above a minimum initial mass of $m_{\star,\min}^{\rm CHE}\sim 2-10~\rm M_\odot$ account for more than half of the total stellar mass following a Salpeter initial mass function. As a result, the UV luminosity functions observed at $z\sim 12-16$ can be reproduced with less extreme values of star formation efficiency and UV luminosity stochastic variability. Our results highlight the potential of CHE in explaining the UV-bright galaxy populations detected by {\it JWST} and call for future work to explore the broader astrophysical implications of CHE and its associated phenomena in the early universe, such as gamma-ray bursts, compact object binaries, and metal enrichment.

\end{abstract}

\keywords{Population III stars (1285) -- Population II stars (1284) -- High-redshift galaxies (734) –- James Webb Space Telescope (2291) –- Early universe (435) -- Stellar evolution (1599)}

\section{Introduction} \label{sec:introduction}
Early results from the {\it JWST} have challenged the standard model of galaxy formation in the high-$z$ universe \citep{Boylan2023}, with the discovery of a surprising abundance of UV bright galaxies at $z\gtrsim 10$ \citep[e.g.,][]{Donnan2023,Donnan2024,Finkelstein2023,Finkelstein2024,Harikane2023,Labbe2023,Adams2024}. Two broad classes of explanations have been discussed, modifications to the underlying $\Lambda$CDM model of cosmological structure formation \citep[e.g.,][]{LiuBromm2022,Shen2024}, or models that vary assumptions in the standard picture of star formation. The second (baryonic) category explores scenarios that either increase the stellar mass/star formation rate in a given dark matter host halo, by boosting the star formation efficiency \citep[SFE, e.g.,][]{Inayoshi2022jwst,Dekel2023} or variability/burstiness of UV emission/star formation \citep[e.g.,][]{Shen2023}, or those that enhance the stellar light-to-mass ratio \citep[e.g.,][]{Donnan2025}. An example for the latter explores a top-heavy initial mass function \citep[IMF; e.g.,][]{Inayoshi2022jwst,Cueto2024,Trinca2024,Ventura2024}, as predicted for the metal-poor environments in the early universe \citep[e.g.,][]{Liu_analytical2024}. {UV emission can also be enhanced when strong radiation-pressure driven outflows remove UV dust attenuation during the early evolution of high-$z$ galaxies \citep[e.g.,][]{Ferrara2024}.}

Invoking a top-heavy IMF or an increased formation efficiency may, however, only boost star formation locally, but suppress it on galactic scales, by triggering disruptive stellar feedback, such as strong supernova (SN) activity \citep[e.g.,][]{Jeong2024}. Here we propose a different (baryonic) solution for the early UV-bright galaxy problem: The stellar light-to-mass ratio is boosted because of a population of stars that undergo chemically homogeneous stellar evolution (CHE), which become more compact and hotter than {stars without CHE}\footnote{{Similar but weaker effects can also be achieved by initial helium enhancement \citep[][]{Katz2024}.}}. Such CHE stage could be realized for rapidly rotating stars, where their structure adapts by triggering strong internal mixing currents \citep[e.g.,][]{Maeder1987, YL2005,Cantiello2007, Yoon2012, MaederMeynet2012, Eldridge2011,
Eldridge2012, Szecsi2015, Szecsi2022, Eldridge2016, Song2016, Mandel2016, deMink2016, Marchant2017, Cui2018, Agui2018, Kubatova2019, Buisson2020, Riley2021, Ghodla2023,
Doroz2024}. Rotation speeds close to the break-up value have been predicted for the first Population~III (Pop~III) stars, formed in the unique metal-free conditions\footnote{Magnetic fields arising from dynamo activity in the primordial protostellar disks could complicate the situation, leading to a possible bifurcation into cases of rapid and slow rotation \citep{Hirano2018}.} of dark matter minihalos \citep{Stacy2011, Stacy2013}. Similarly, metal-poor Population~II (Pop~II) stars may be able to retain their natal rotation speeds, because of their significantly reduced mass loss due to weakened radiatively-driven stellar winds \citep[e.g.,][]{Vink2001}. {Recent direct observations of select O-type stars in nearby metal-poor dwarf galaxies suggest that stellar winds may be even weaker than what was previously predicted by theory \citep{Sander2017,Sander2018,Mata2024, Telford2024}.}

The impact of {a CHE-boost in the ionizing radiation of H, He, and possibly He$^{+}$} from Pop~III stars on the early stages of reionization was explored in a previous study \citep{Sibony2022}, placing constraints on the fraction of Pop~III stars that undergo CHE \citep[see also e.g.][and Liu et al. in prep.]{Ghodla2023}. The reionization modeling, however, is subject to the uncertain escape fraction of ionizing radiation from the galaxy host halos, whereas any constraints from the non-ionizing UV continuum are more direct indications of the underlying stellar populations. We note that there have been recent indications for the metal enrichment from rapidly rotating stars in the first galaxies, inferred from their chemical abundance patterns with deep {\it JWST} spectroscopy. Prominent among them are high nitrogen-to-oxgen abundance ratios, e.g., in GN-z11 and RXCJ2248-ID \citep{Bunker2023,Topping2024}, which may be explained with models of rapidly rotating, massive stars \citep[e.g.,][]{Tsiatsiou2024, Nandal2024}. Another nucleosynthetic signature of rapidly rotating stars is the proposed boost in carbon production \citep{Liu_winds2021,Jeena2023}, providing a possible explanation for strong carbon emission lines detected in $z\gtrsim 12$ galaxies \citep[e.g.,][]{DEugenio2024} and carbon-enhanced extremely metal-poor stars in the Milky Way \citep[e.g.,][]{Zepeda2023}. {As the extreme outcome of rotational mixing, CHE leads to the maximal enhancement of carbon production \citep{Jeena2023}, although it is unclear whether nitrogen enhancement can be achieved by stars with CHE, which presumably depends on how the CHE phase starts and ends.} 
We will therefore explore in the following whether a rapidly rotating subset of stars under CHE within the first galaxies can account for the emerging {\it JWST} phenomenology of UV bright galaxies in the first few hundred million years of cosmic history.  
  
\section{Chemically Homogeneous Evolution} \label{sec:CHE}

\subsection{Stellar Evolution}
\label{sec:CHE_Struct}

\begin{figure*}[htbp]
    \centering
    \includegraphics[width=0.9\textwidth]{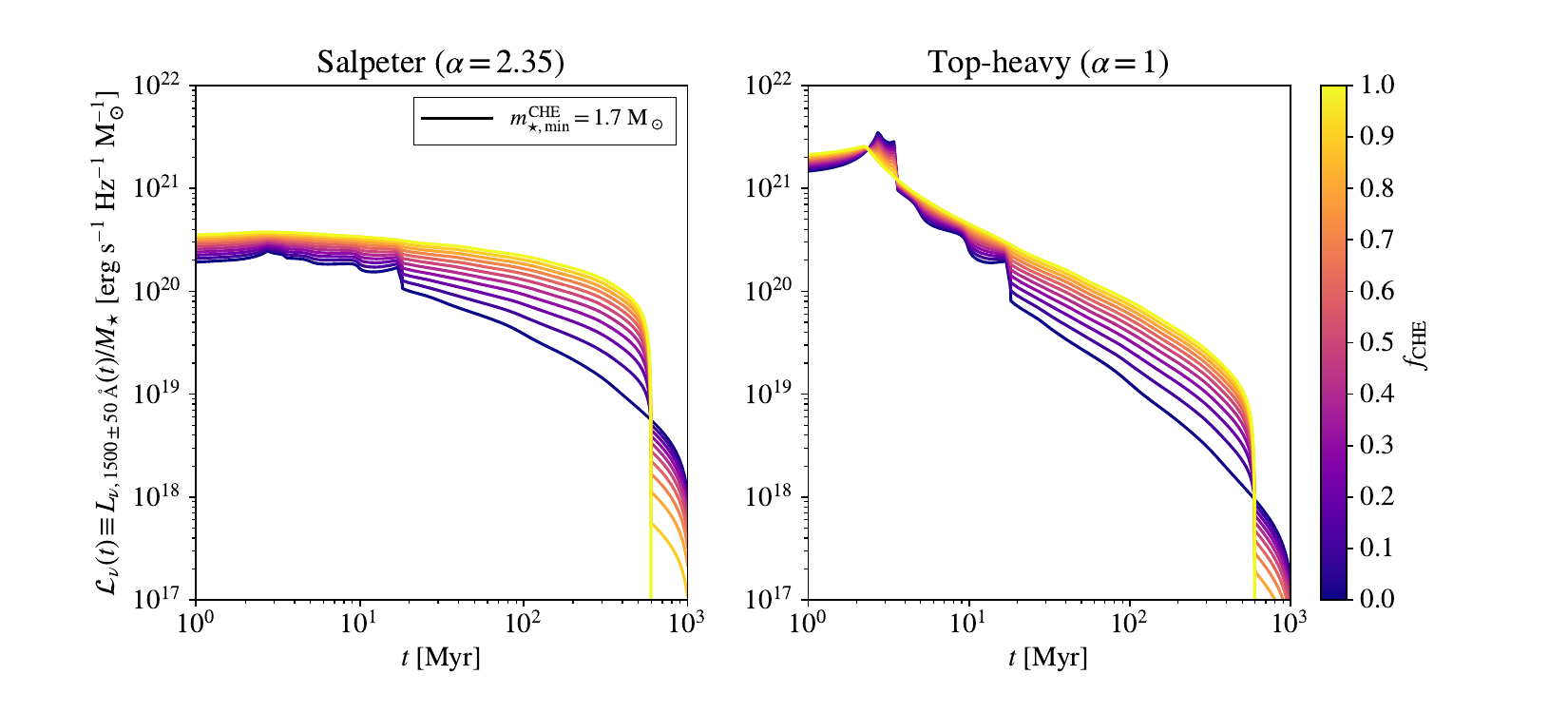}%
    \caption{Rest-frame UV light curve: evolution of UV luminosity normalized by the total stellar mass for a single-age stellar population. The left (right) panel shows the results for a Salpeter (top-heavy) IMF, $dN/dm_{\star}\propto m_{\star}^{-\alpha}$, with a slope $\alpha=2.35$ (1) and a mass range $m_{\star}\in [1.7,300]~\rm M_\odot$ for both {non-rotating} and CHE stars. The mass fraction of CHE stars $f_{\rm CHE}$ is varied between 0 and 1, with lighter colors denoting larger $f_{\rm CHE}$.}
    \label{fig:light_curve}
\end{figure*}

The stellar evolution models for stars evolving homogeneously are based on $n = 3$ polytropes with a time-varying mass fraction of hydrogen, as detailed in \citet{Sibony2022}. In that work Pop~III stars with initial masses ($m_{\star}$) between 9 and 300~M$_\odot$ were computed. Here we complement our previous grid by adding tracks for lower initial mass stars with masses between 1.7 and 7~M$_\odot$. 
As a comparison sample, we use the stellar tracks for non-rotating Pop~III stars in the same mass range ($1.7-300~\rm M_\odot$) from \citet{Murphy2021}. Here, we only consider {initially} metal-free stellar evolution models for simplicity, while freely varying the IMF parameters (see Sec.~\ref{sec:CHE_Lum} below). {We focus on the Main-Sequence (MS) phase where most UV photons are produced and ignore mass loss from stellar winds, which is minor for metal-poor stars \citep[e.g.,][]{Hainich2015,Kubatova2019}. For metal-rich massive stars where winds are strong, the UV emission will be overestimated in our models. This renders our results as optimistic estimates. In fact, for non-rotating stars,} the difference in rest-frame UV ($\lambda\sim 1500~\rm \AA$) luminosity made by changing the {initial} metallicity from $Z=0$ to $Z=0.02$ under a fixed \citet{Kroupa2001} IMF is small (within a factor of $\sim 2$) for a typical star formation timescale~$\sim 100$~Myr, according to the Yggdrasil stellar population synthesis models \citep[]{Zackrisson2011}. {For CHE stars, \citet{Kubatova2019} found that changing the wind parameters (mass loss rate and clumping factor) has negligible effects on the UV spectrum around $\lambda\sim 1000-3000\ \rm\AA$ during MS with helium mass fractions $Y\lesssim 0.98$ for $m_\star\sim 20-130\ \rm M_\odot$ in the metal-poor regime ($Z=0.0002$), although the impact of winds can be stronger at higher metallicities. However, the impact is expected to be smaller than the case of non-rotating stars since the strong surface enhancement of helium is the dominant factor in CHE, as implied by the smaller effects of metallicity on the properties of pure helium stars \citep[see, e.g., figs.~5 and 6 in][]{Iorio2023}. }

Homogeneous evolution requires that the timescale for chemical mixing is much smaller than the nuclear timescale during the MS phase. The mixing timescale scales as $R^2/D_{\rm mix}$, where $R$ is the stellar radius, and $D_{\rm mix}$ the diffusion coefficient describing the transport of the chemical elements. \citet[][see their paragraph~3.1]{MM2000} showed that the ratio of the mixing timescale to MS lifetime approximately varies as $1/m_{\star}$ for stars more massive than $15~\rm M_{\odot}$. This indicates that when the mass decreases, the mixing timescale increases faster than the MS lifetime. Thus, we can expect to reach a lower initial mass limit, $m_{\star,\min}^{\rm CHE}$, below which homogeneous evolution will not be obtained. The precise limit depends on the initial rotation and metallicity, with lower values when the initial rotation increases and when the metallicity decreases {\citep[see, e.g., fig~4 in][]{Brott2011}, which is due at least in part to the fact that at lower metallicities, stars are more compact}.

This limit also depends on the physics used for chemical mixing and for the transport of angular momentum. These two transport mechanisms are intimately related since the gradients of chemical composition impact angular momentum transfer, which in turn influences the transport of the chemical elements. For this work, since we do not use a peculiar physical mechanism to obtain the homogeneous tracks, we simply consider the value of this limit as a free parameter. Most of the 
stellar models computed so far have focused on the massive stellar mass range
down to lower masses between {4 and 10~M$_\odot$ \cite[see, e.g.,][]{Yoon2006,Brott2011,Szecsi2015,Riley2021}}, so we may wonder if it is justified here to extend this limit further. As indicated above, the limit depends on the physics of mixing which is still not yet fully
understood. On the other hand, fast rotation is likely not the only process that may lead to homogeneous stars. For instance, grazing collisions have been found to result in nearly completely mixed stars producing blue stragglers, as observed in globular clusters \citep[e.g.,][]{Benz1987}.

\subsection{{UV Emission and stellar populations}}
\label{sec:CHE_Lum}

As in \citet{Sibony2022}, we consider stars to radiate as black bodies and compute their average spectral luminosity $L_{\star,\nu}(t,m_{\star})$ in a $100$~\r{A} window around the typical rest-frame UV wavelength of $1500$~\r{A} along their evolutionary tracks. {We have verified using detailed stellar atmosphere models computed by the code \textsc{tlusty} \citep{Hubeny1988,Hubeny2021} that the relative errors in $L_{\star,\nu}(t,m_{\star})$ introduced by the back-body assumption are within $30$\% for most CHE and non-rotating stars with $m_{\star}\sim 3-300\rm~M_\odot$, surface gravities $g\sim 10^{4.1-5.1}~\rm cm~s^{-2}$, and effective temperatures $T_{\rm eff}\sim 10^{4.3}-10^{5}~\rm K$, that dominate the overall rest-frame UV emission.} {We also compare our results with those from \citet{Kubatova2019} for CHE stars with $Z=0.0002$ and $m_{\star}\sim 20-130\ \rm M_\odot$, whose spectra are calculated using the atmosphere code \textsc{powr} \citep[e.g.,][]{Sander2015} from the evolutionary tracks computed by \citet{Szecsi2015}. The discrepancy is also within 30\% during MS ($Y\lesssim 0.98$), mainly resulting from the different evolutionary tracks. When $T_{\rm eff}$ is controlled, the difference caused by the black-body assumption alone is within 7\%. It is shown below that our conclusions are robust despite these uncertainties. }

We consider a mixed stellar population containing both CHE and non-rotating stars as a single-age starburst. This population is defined by two parameters, the mass fraction $f_\mathrm{CHE}\in[0,1]$ of CHE stars, and the minimal stellar mass $m_{\star,\min}^{\rm CHE}\in~[1.7,20]\rm~M_\odot$ for CHE to occur. For a given choice of $(f_\mathrm{CHE},m_{\star,\min}^{\rm CHE})$, we first generate two stellar populations following a power-law IMF, $dN/dm_{\star}\propto m_{\star}^{-\alpha}$, with a slope of $\alpha=2.35$ \citep[]{Salpeter1955} or $\alpha=1$ (top-heavy). {The population of CHE stars is generated between $m_{\star,\min}^{\rm CHE}$ and $m_{\star,\max}=~300\rm~M_\odot$, while non-rotating stars span a fixed mass range from $m_{\star,\min}=1.7~\rm M_\odot$ to $m_{\star,\max}=300\ \rm M_\odot$.} For each population, we compute the total luminosity $L_\nu(t)$ by summing up the $L_{\star,\nu}(t,m_{\star})$ for all stars that are still alive at time $t$, up to $t=1~$Gyr, which is then normalized by the total initial stellar mass $M_{\star}$ as $\mathcal{L}_{\nu}(t)\equiv L_{\nu}(t)/M_{\star}$. Finally, we combine the CHE and non-rotating populations so that the mass fraction of CHE stars is $f_\mathrm{CHE}$. For illustration, Figure~\ref{fig:light_curve} shows the resulting rest-frame UV light curves for $m_{\star,\min}^{\rm CHE}=~1.7\rm~ M_\odot$ and varying $f_\mathrm{CHE}$. 

Note that in the aforementioned default model, the overall IMF of the mixed population is no longer a strict power-law when $f_{\rm CHE}>0$ and $m_{\star,\min}^{\rm CHE}>m_{\star,\min}=1.7~\rm M_\odot$ as the CHE component is more top-heavy than the non-rotating component. An alternative approach is to turn a fraction of the stars above $m_{\star,\min}^{\rm CHE}$ into CHE stars while keeping the overall IMF fixed to the chosen power-law form. In this case, the mass fraction of CHE stars has an upper limit $f_{\rm CHE}^{\max}=[\int_{m_{\star,\min}^{\rm CHE}}^{m_{\star,\max}}m_{\star}(dN/dm_{\star})dm_{\star}]/[\int_{m_{\star,\min}}^{m_{\star,\max}}m_{\star}(dN/dm_{\star})dm_{\star}]$ as a function of $m_{\star,\min}^{\rm CHE}$. Although we do not model this case explicitly, its outcomes can be estimated from the results of the default model via $f_{\rm CHE}^{\max}$, which do not change our conclusions made with the default model.

\begin{figure*}
    \centering
    \includegraphics[width=0.9\textwidth]{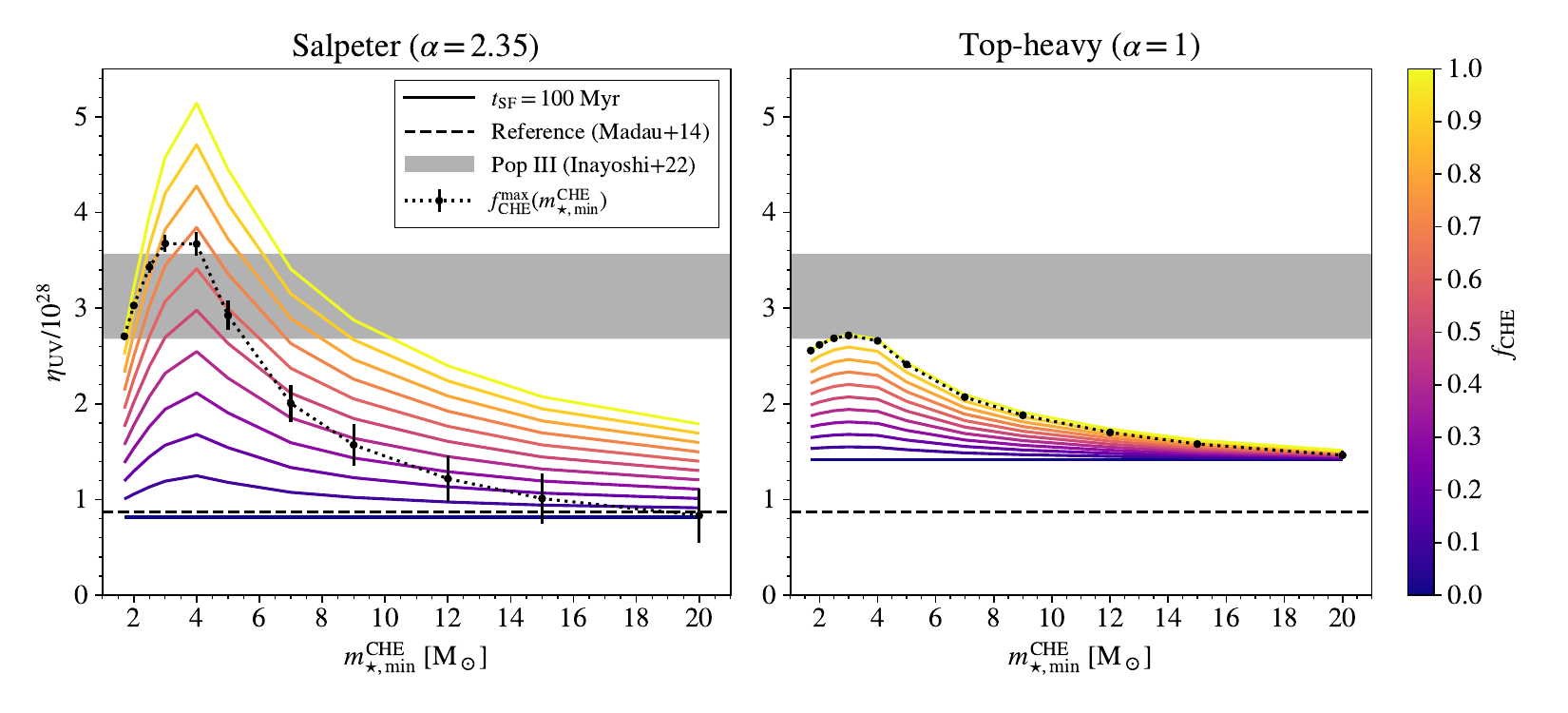}
    \caption{Rest-frame UV luminosity per unit SFR $\eta_{\rm UV}$ for a top-hat star formation history within a timescale $t_{\rm SF}=100~\rm Myr$. As in Fig.~\ref{fig:light_curve}, the left and right panel show the results for Salpeter and top-heavy IMFs, with $\alpha=2.35$ and $1$, respectively. While the mass range of {non-rotating} stars is fixed as $m_{\star}\in [1.7,300]~\rm M_\odot$, we plot $\eta_{\rm UV}$ as a function of $m_{\star,\min}^{\rm CHE}$ for $f_{\rm CHE}\in [0, 1]$, where $f_{\rm CHE}$ increases for lighter colors (from bottom to top). The dotted curve with errorbars shows the maximum $\eta_{\rm UV}$ in the alternative population model with a fixed overall IMF (corresponding to the maximum CHE mass fraction $f_{\rm CHE}^{\max}$), which is estimated to be in between $\eta_{\rm UV}(f_{\rm CHE}=f_{\rm CHE}^{\max})$ and $f_{\rm CHE}^{\max}\eta_{\rm UV}(f_{\rm CHE}=1)$. 
    We also show the reference value $\eta_{\rm UV}\simeq 8.70~\times~10^{27}$ from \citet[][]{Madau2014} with the dashed line, which is very close to our result for $f_{\rm CHE}=0$ given $\alpha=2.35$. The shaded regions show the results for Pop~III stars with extremely top-heavy IMFs from \citet{Inayoshi2022jwst}.}
    \label{fig:kuv_mmin}
\end{figure*}

\section{Impact on First Galaxy UV Emission}
\label{sec:UVLF}

To predict the UV magnitudes of high-$z$ galaxies and make direct comparisons with observations, we focus on the UV emission efficiency $\eta_{\rm UV}\equiv \kappa_{\rm UV}^{-1}$ defined as the inverse of the conversion factor $\kappa_{\rm UV}$ between star formation rate (SFR) and rest-frame UV luminosity $L_\nu(\rm UV)$:
\begin{equation}
    {\rm SFR/(M_\odot~yr^{-1})}=\kappa_{\rm UV}L_\nu(\rm UV)/(erg~s^{-1}~Hz^{-1})~.
\end{equation}
{A reference value from local calibrations adopted by \citet{Madau2014} is $\eta_{\rm UV}\simeq8.70\times10^{27}$, corresponding to $\kappa_{\rm UV}=1.15\times~10^{-28}$.} Given the single-age starburst results (see Fig.~\ref{fig:light_curve}), we calculate $\eta_{\rm UV}$ as a function of the CHE fraction $f_{\rm CHE}$ and minimum mass $m_{\star,\min}^{\rm CHE}$ under different IMFs by integrating the stellar mass-normalized light curve $\mathcal{L}_{\nu}(t)$ over a simple top-hat star formation history with a constant SFR for a period of $t_{\rm SF}$:
\begin{equation}
    \eta_{\rm UV}=\int_{0}^{t_{\rm SF}}\mathcal{L}_{\nu}(t)dt\times \rm M_\odot~yr^{-1}/(erg~s^{-1}~Hz^{-1})~.\label{kuv}
\end{equation}
We adopt $t_{\rm SF}=100$~Myr by default (see Appendix~\ref{sec:kuv_tsf} for the results for $t_{\rm SF}=10$ and 1000~Myr) as the typical timescale of star formation probed by rest-frame UV emission, which is also close to the halo dynamical timescale and baryon cycle timescale at $z\sim 10$ \citep[e.g.,][]{Angles-Alcazar2017,FloresVelazquez2021,Tacchella2020}. 
We have ignored the contribution of nebular emission and dust attenuation. {As discussed in Appendix~\ref{sec:nebular}, for moderately metal-enriched ($Z\gtrsim 0.0004$) luminous high-$z$ galaxies observable by \textit{JWST}, the enhancement of rest-frame UV emission by CHE with $t_{\rm SF}\gtrsim 10$~Myr can be underestimated by up to a factor of $\sim 2$ in the absence of nebular emission, which is also expected to be enhanced for CHE stars due to their strong ionizing power \citep{Kubatova2019,Sibony2022}, and the effect is stronger for more metal-poor systems \citep{Zackrisson2011}. Therefore, our results without nebular contributions should be regarded as conservative estimates.}
The effect of dust attenuation can be effectively absorbed into the UV variability parameter \citep[see below and][]{Shen2023}. 

%


The results are shown in Fig.~\ref{fig:kuv_mmin}. We obtain $\eta_{\rm UV}\simeq~8.17~\times~10^{27}$ in the canonical case of Salpeter IMF without CHE ($f_{\rm CHE}=0$), which is very close to the reference value $\eta_{\rm UV}\simeq 8.70~\times~10^{27}$ \citep{Madau2014}. This further justifies our choice of $t_{\rm SF}=100$~Myr (as well as $m_{\star,\min}=1.7~\rm M_\odot$). Interestingly, the enhancement of UV emission by CHE peaks at $m_{\star,\min}^{\rm CHE}=4~(3)~\rm M_\odot$ for the Salpeter (top-heavy) IMF, reaching $\eta_{\rm UV}\sim 5.14~(2.73)~\times~10^{28}$ for $f_{\rm CHE}=1$, which is higher than the reference value by a factor of $\sim 6$ (3). Here, $\eta_{\rm UV}$ does not increase monotonically with $m_{\star,\min}^{\rm CHE}$ because the (lifetime-integrated) rest-frame UV emission {of CHE stars has a peak around $m_\star\sim 2~\rm M_\odot$ and decreases for more massive stars, dropping even below the values for non-rotating stars at $m_\star\gtrsim 100~\rm M_\odot$. 
Here, massive stars} become too hot under CHE, producing primarily more energetic (ionizing) photons \citep[Liu et al. in prep.]{Sibony2022}. 
For the same reason, the effect of CHE is generally stronger at $m_{\star,\min}^{\rm CHE}\lesssim 10~\rm M_\odot$ when $\alpha$ or $t_{\rm SF}$ increase (see Appendix~\ref{sec:kuv_tsf}) or with a smaller IMF upper mass limit $m_{\star,\max}$. Without CHE, we obtain $\eta_{\rm UV}=1.42~\times~10^{28}$ under our top-heavy IMF (that extends down to $m_{\star,\min}=1.7~\rm M_\odot$ with $\alpha=1$), moderately increased by $\sim 74$ (63)\% compared with the Salpeter (reference) case. It is shown in \citet[see their table.~1]{Inayoshi2022jwst} that considering extremely top-heavy IMFs ($m_{\star}\gtrsim 10~\rm M_\odot$) for Pop~III stars further increases the UV emission efficiency to $\eta_{\rm UV}\sim 2.68-3.57~\times~10^{28}$ \citep[see also][]{BKL2001}. Comparable and higher values of $\eta_{\rm UV}$ can be achieved by CHE with $f_{\rm CHE}\gtrsim 0.5$ and $m_{\star,\min}^{\rm CHE}\sim 2-10~\rm M_\odot$ under a Salpeter IMF. 

Given the results of $\eta_{\rm UV}(f_{\rm CHE})$ in the default model, the maximum UV efficiency in the alternative case of a fixed overall IMF should be in between $\eta_{\rm UV}(f_{\rm CHE}=f_{\rm CHE}^{\max})$ (with optimistic contributions from non-rotating stars below $m_{\star,\min}^{\rm CHE}$) and $f_{\rm CHE}^{\max}\eta_{\rm UV}(f_{\rm CHE}=1)$ (ignoring non-rotating stars below $m_{\star,\min}^{\rm CHE}$). The resulting values are very close to $\eta_{\rm UV}(f_{\rm CHE}=1)$ for the top-heavy IMF ($\alpha=1$) as $f_{\rm CHE}^{\max}>0.94$. For the Salpeter IMF with $f_{\rm CHE}^{\max}\sim 0.3-1$, the maximum UV emission efficiency peaks around $m_{\star,\min}^{\rm CHE}\sim 3-4~\rm M_\odot$ given $f_{\rm CHE}^{\max}\sim 0.7-0.8$, reaching $\eta_{\rm UV}\sim 3.6-3.8~\times~10^{28}$, which is still a factor of $\sim 4$ larger than the reference value. 

Once $\eta_{\rm UV}$ is known, we derive the UV luminosity function (UVLF) using the model in \citet{Shen2023}. First, we calculate the median UV magnitude $M_{\rm UV}$ as a function of halo mass $M_{\rm h}$ from the median ${\rm SFR}=~\epsilon_{\star}f_{\rm b}\dot{M}_{\rm h}$ estimated by the median halo accretion rate \citep{Fakhouri2010}
\begin{equation}
\begin{split}
    \dot{M}_{\rm h}&\simeq 25.3~{\rm M_\odot~yr^{-1}}\left(\frac{M_{\rm h}}{10^{12} {\rm M_\odot}}\right)^{1.1}\\
    &\times (1+1.65z)\sqrt{\Omega_{\rm m}(1+z)^{3}+\Omega_{\Lambda}}~,\label{mdot_halo}
\end{split}
\end{equation}
and SFE \citep[e.g.,][]{Tacchella2018,Harikane2022}
\begin{equation}
    \epsilon_{\star}=\frac{2\epsilon_0}{(M_{\rm h}/M_0)^{-\gamma}+(M_{\rm h}/M_0)^{\xi}}~.\label{sfe}
\end{equation}
Here $f_{\rm b}\equiv\Omega_{\rm b}/\Omega_{\rm m}\simeq 0.16$ is the cosmic baryon fraction, and we adopt $\epsilon_0=0.1$, $\gamma=0.6$, and $\xi=0.5$ based on \citet[for the normalization and low-mass slope, matching the ${\rm SFR}$-$M_{\rm h}$ relation at $z=7$]{Behroozi2019} and \citet[for the high-mass slope]{Harikane2022}. Next, this median $M_{\rm UV}$-$M_{\rm h}$ relation is combined with the halo mass function\footnote{We use the \textsc{hmf} package \citep{Murray2013} to calculate the halo mass function with the \citet{Tinker2010} model and real-space top-hat window function for a flat $\Lambda$CDM cosmology. The relevant cosmological parameters are based on \citet{PlanckCollaboration2020} results: $H_0=67.66~\rm km~s^{-1}~Mpc^{-1}$, $\Omega_{\rm m}=0.30966$, $n_{\rm s}=0.9667$, $\sigma_8=0.8159$, and $\Omega_{\rm b}=0.04897$.} $dn/d\log M_{\rm h}$ to derive the \textit{median} UVLF $dn_{\rm median}/dM_{\rm UV}=(dn/d\log M_{\rm h})|\frac{d\log M_{\rm h}}{dM_{\rm UV}}|$, which is then convolved with a Gaussian kernel of width $\sigma_{\rm UV}$ (mag) to predict the observable UVLF $\Phi\equiv dn/dM_{\rm UV}$. The last step captures the stochastic scatter in the scaling relations used above into the UV variability parameter $\sigma_{\rm UV}$. 

\begin{figure}
    \centering
    \includegraphics[width=\linewidth]{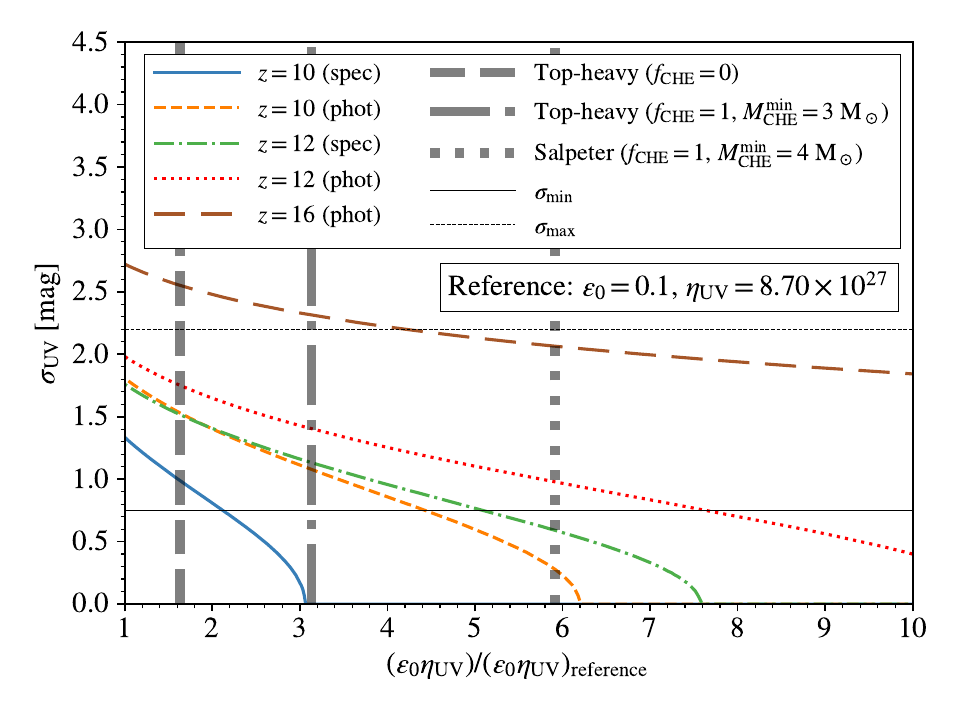}
    \caption{Constraints on UV emission efficiency and variability from {\it JWST} data. The colored curves show the values of $\epsilon_0 \eta_{\rm UV}$ and $\sigma_{\rm UV}$ required to reproduce $\log(\Phi(M_{\rm UV}=~-20.5)~[\rm mag^{-1}~cMpc^{-3}])=-5$ and $-5.3$ at $z=10$ (solid) and 12 (dash-dotted) based on spectroscopy, and $\log(\Phi(M_{\rm UV}=-20.5)~[\rm mag^{-1}~cMpc^{-3}])=-4.5$, $-5$, and $-5.2$ at $z=10$ (dashed), 12 (dotted), and 16 (long dashed) from photometry. Three representative cases of different $\eta_{\rm UV}$ with fixed $\epsilon_0=0.1$ are shown by the thick vertical lines: the top-heavy models with no CHE ($f_{\rm CHE}=0$, dashed) and optimal CHE ($f_{\rm CHE}=1$, $m_{\star,\min}^{\rm CHE}=3~\rm M_{\odot}$, dash-dotted) and the Salpeter model with optimal CHE ($f_{\rm CHE}=1$, $m_{\star,\min}^{\rm CHE}=4~\rm M_{\odot}$, dotted). The thin solid and dashed horizontal lines show the minimum and maximum UV variability estimated by \citet[]{Shen2023}.}
    \label{fig:sigma_k}
\end{figure}

\begin{figure}[htbp]
    \centering
    \includegraphics[width=\linewidth]{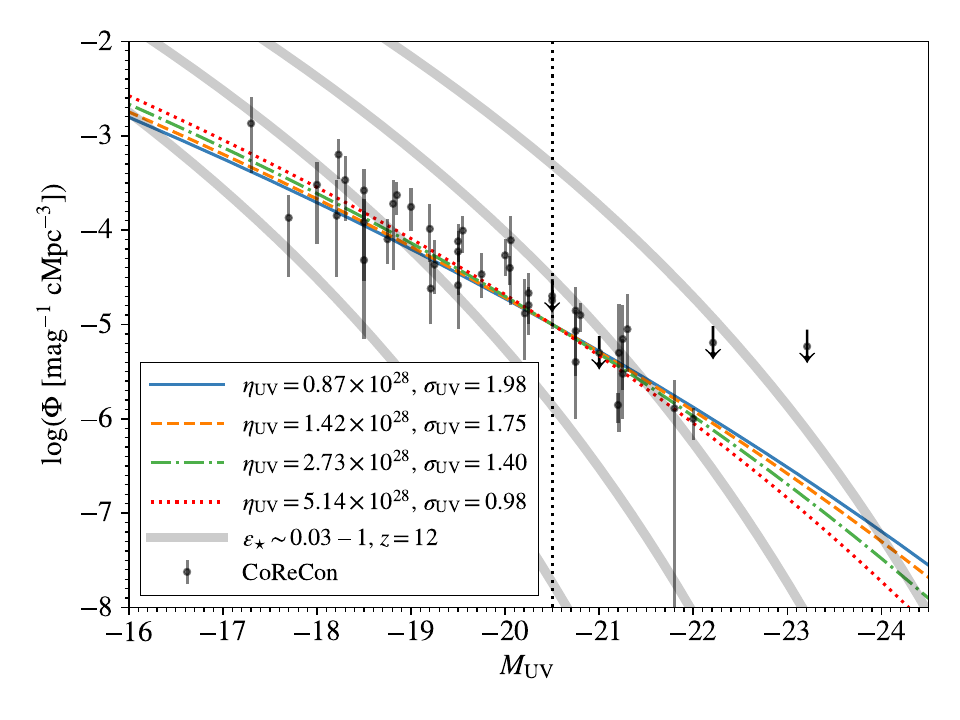}
    \caption{UVLF for select degenerate models that meet the photometric constraints from {\it JWST} at $z=12$, i.e., $\log(\Phi(M_{\rm UV}=-20.5)~[\rm mag^{-1}~cMpc^{-3}])=-5$. These include the reference case with $\eta_{\rm UV}\simeq 8.70~\times~10^{27}$ (solid) and the three representative models with enhanced UV emission from CHE and/or a top-heavy IMF considered in Fig.~\ref{fig:sigma_k}. For comparison, the observational data for $z\sim 11-13$ \citep{Naidu2022two,Bouwens2023,Harikane2023,Perez-Gonzalez2023,Adams2024,Casey2024,Donnan2023,Donnan2024,Finkelstein2024,McLeod2024,Robertson2024,Willott2024} are shown using \textsc{CoReCon} \citep{Garaldi2023}, among which the data points with arrows are upper limits. We also plot the results for constant SFE $\epsilon_{\star}=0.03$, 0.1, 0.3, and 1 (from bottom to top) without UV variability as the thick solid lines.}
    \label{fig:uvlf_comp}
\end{figure}

As illustrated in \citet[][see their fig.~3]{Shen2023}, the UVLFs inferred from {\it JWST} observations place constraints on $\epsilon_0 \eta_{\rm UV}$ and $\sigma_{\rm UV}$, which can be estimated by considering the value of $\Phi$ at a typical magnitude $M_{\rm UV}=-20.5$. Here, we require $\log(\Phi(M_{\rm UV}=~-20.5)~[\rm mag^{-1}~cMpc^{-3}])=-5$ and $-5.3$ at $z=10$ and 12 according to spectroscopic observations, and $\log(\Phi(M_{\rm UV}=-20.5)\rm~[mag^{-1}~cMpc^{-3}])=-4.5$, $-5$, and $-5.2$ at $z=10$, 12, and 16 from photometric results, 
using the observational data compiled by the \texttt{python} package \textsc{CoReCon}\footnote{\url{https://corecon.readthedocs.io/en/latest/index.html}} 
\citep{Garaldi2023}. The corresponding values of $\epsilon_0 \eta_{\rm UV}$ and $\sigma_{\rm UV}$ are shown in Fig.~\ref{fig:sigma_k}, where larger $\epsilon_0 \eta_{\rm UV}$ and/or $\sigma_{\rm UV}$ are required to explain observations at higher $z$, and a smaller $\epsilon_0 \eta_{\rm UV}$ can be compensated with a larger $\sigma_\mathrm{UV}$. 

To appreciate the trend and the degeneracy, we focus on 3 representative cases with boosted UV emission efficiency and fixed $\epsilon_0=0.1$: the top-heavy ($\alpha=1$) models with no CHE ($f_{\rm CHE}=0$) and optimal CHE ($f_{\rm CHE}=1$, $m_{\star,\min}^{\rm CHE}=4~\rm M_{\odot}$), and the Salpeter ($\alpha=2.35$) model with optimal CHE ($f_{\rm CHE}=1$, $m_{\star,\min}^{\rm CHE}=3~\rm M_{\odot}$). The corresponding UV variability parameters inferred from the photometric constraints at $z\sim 12$ (16) are $\sigma_{\rm UV}=1.75$, 1.40, and 0.98 (2.55, 2.31, and 2.06), respectively, while the reference value $\eta_{\rm UV}\simeq 8.70~\times~10^{27}$ requires $\sigma_{\rm UV}=1.98$ (2.72). Clearly, with enhanced $\eta_{\rm UV}$ from CHE, smaller (less extreme) values of $\epsilon_0$ and $\sigma_{\rm UV}$ are required to explain {\it JWST} results. In particular, only for the Salpeter model with optimal CHE ($f_{\rm CHE}=1$, $m_{\star,\min}^{\rm CHE}=3~\rm M_\odot$, $\alpha=2.35$) the $\sigma_{\rm UV}$ needed at $z\sim 16$ remains below the maximum value $2.2$ estimated by \citet{Shen2023}. Fig.~\ref{fig:uvlf_comp} shows the UVLFs of the (degenerate) models at $z\sim 12$, which all agree well with current observations, but differ slightly in shape: The evolution of $\Phi$ with $M_{\rm UV}$ is stronger for larger (smaller) $\epsilon_0 \eta_{\rm UV}$ ($\sigma_{\rm UV}$), especially at the luminous end. Therefore, high precision measurements of UVLFs (in larger survey volumes) can potentially break the degeneracy between $\epsilon_0 \eta_{\rm UV}$ and $\sigma_{\rm UV}$. Additional constraints from galaxy clustering properties are also useful \citep{Munoz2023,Sun2024}.

\section{Summary and Discussion}
We explore the contribution of rapidly rotating stars under chemically homogeneous evolution (CHE) to the rest-frame UV ($\lambda\sim 1500~\rm\AA$) emission of the first galaxies. By incorporating CHE into a simple population synthesis model under the assumptions of \textit{black-body spectrum} and \textit{negligible nebular emission}, we predict the UV emission efficiency $\eta_{\rm UV}\propto L_{\nu}(\rm UV)/SFR$ 
as a function of the mass fraction $f_{\rm CHE}$ and minimum mass $m_{\star,\min}^{\rm CHE}$ of CHE stars with two power-law initial mass function (IMF) models ($dN/dm_{\star}\propto m_{\star}^{-\alpha}$, $\alpha=2.35$ and~1, $m_{\star}<300~\rm M_\odot$). We then adopt the model in \citet{Shen2023} to calculate the relevant UV luminosity function (UVLF). 

We find that CHE can significantly enhance the rest-frame UV emission of star-forming galaxies, as efficient as alternative scenarios of enhanced star formation efficiency and top-heavy IMFs. Compared with the canonical value from local calibrations \citep{Madau2014}, $\eta_{\rm UV}$ is increased by a factor of $\sim 3-6$ for $f_{\rm CHE}\gtrsim 0.5$ and $m_{\star,\min}^{\rm CHE}\sim 2-10~\rm M_\odot$ with a Salpeter IMF ($\alpha=2.35$) at the typical star-formation timescale $t_{\rm SF}\sim 100$~Myr. The effect of CHE is generally stronger for larger $t_{\rm SF}$ or IMF slope $\alpha$. 
The enhanced UV emission from CHE can reproduce the UVLFs inferred from observations at $z\sim 12-16$ with less extreme values of star formation efficiency (SFE) and stochastic variability of UV luminosity, within the range of current theoretical predictions. {Taking into account nebular emission can further enhance the effect of CHE by up to a factor of~$\sim 2$ for $Z\gtrsim 0.0004$ (see Appendix~\ref{sec:nebular}). }

{Moreover, compared with top-heavy IMFs, a potential advantage of the CHE scenario is that the enhancement of SN feedback is likely weaker, making it less vulnerable to the global suppression of star formation by such feedback \citep{Jeong2024}. 
We estimate the (IMF-averaged) SN energy per unit stellar mass $\mathcal{E}_{\rm SN}$ for our pure CHE and non-rotating populations using the SN models in \citet{Goswami2022}, assuming that the SN explosion of a star undergoing CHE is identical to that of the (pre-SN) helium core with the same mass (ignoring mass loss)\footnote{{We have ignored two competing effects possibly associated with CHE: the enhancement of SN energy by fast rotation \citep[e.g.,][]{Burrows2007,MaederMeynet2012,Chen2015,Fujibayashi2023}, and the reduction of core mass by strong Wolf-Rayet-like winds during post MS or near the end of MS \citep[e.g.,][]{Yoon2012,Kubatova2019,Riley2021,Umeda2024}. Taking them into account in a detailed investigation of the SN feedback from CHE stars is deferred to future work.}}. In the metal-poor regime ($Z\lesssim 0.01$) of high-$z$ galaxies, $\mathcal{E}_{\rm SN}$ is enhanced by a factor of $\sim 25-34$ in the top-heavy ($\alpha=1$) model of non-rotating stars compared with the canonical case of Salpeter IMF ($\alpha=2.35$). On the other hand, in the optimal CHE model with $\alpha=2.35$ and $m_{\star,\min}^{\rm CHE}=4~\rm M_\odot$, $\mathcal{E}_{\rm SN}$ is only boosted by a factor of $\sim 8-11$, comparable to the increase of $\eta_{\rm UV}$. 
This means that the effects of CHE on UV emission and SN feedback can be similar to those of increasing the SFE. However, unlike the latter, CHE is not limited by the baryon mass budget.} 

{Besides, stellar UV spectra become harder under CHE \citep[see also][Liu et al. in prep.]{Sibony2022} 
which may play a role in the decrease of UV continuum slope $\beta$ (from the fit $L_{\lambda}\propto \lambda^{\beta}$ for $\lambda\sim 1270-2600\ \rm\AA$) with redshift and extremely blue galaxies with $\beta\lesssim -2.6$ in observations \citep[e.g.,][]{Dottorini2024}. In our top-heavy (Salpeter) IMF models, for CHE stars with $m_{\star,\min}^{\rm CHE}=1.7\ \rm M_\odot$, we obtain $\beta\sim -3$ ($-2.4$) and $-2.3$ ($-2$) at 10 and 100~Myr after the starburst, respectively, while non-rotating stars give larger values, $\beta\sim -2.7$ ($-2.2$) and $-1.2$ ($-1$), under the same condition\footnote{{Note that these results do not consider nebular enmission and treat stars as black bodies. According to the \textsc{tlusty} stellar atmosphere models, the spectral shape of UV continuum at $\lambda\sim 1270-2600\ \rm\AA$ does not deviate much from that of black bodies for hot stars. Besides, the nebular contributions can be small in extremely blue galaxies due to high escape fractions and/or bursty star formation \citep{Dottorini2024}}.}.}


{Although this work focuses on the UV continuum, stellar and nebular lines are also affected by CHE and fast rotation in general \citep[e.g.,][]{Kubatova2019,Newman2025}.} In addition to boosting the UV emission \citep[as well as ionizing power, see, e.g.,][Liu et al. in prep.]{Sibony2022} 
of galaxies, populations of chemically homogeneous stars may have consequences for the nature of transients and for stellar nucleosynthesis. If homogeneous evolution is triggered by fast rotation, those stars (at least the massive ones) may produce long soft gamma ray bursts (GRBs), via the collapsar mechanism \citep[e.g.,][]{YL2005,Yoon2006, Cantiello2007, Yoon2012}. With the extension to high redshifts considered here, GRBs may then also originate in the first galaxies \citep[e.g.,][]{Bromm2006}, and their associated bright afterglows could act as background sources to probe the early intergalactic medium with absorption spectroscopy \citep{Wang2012}. The latter would be an ideal target for the upcoming suite of extremely large telescopes (ELTs). 
{CHE can play important roles in post-MS and binary stellar evolution, regulating the properties of evolved stars and binary products, such as Wolf-Rayet stars, X-ray binaries, and binary black hole mergers \citep[e.g.,][]{Eldridge2012,Yoon2012,Mandel2016, deMink2016, Marchant2017,Kubatova2019, Buisson2020, Riley2021,Umeda2024,Dall'Amico2025}}. Moreover, those stars, if they lose mass, may enrich their surroundings in helium, hydrogen-burning products, and possibly also some helium-burning products, while being depleted in lithium. This material could furthermore exhibit strong carbon over-abundances, thus leading to the formation of next-generation carbon-enhanced metal-poor stars \citep{Liu_winds2021,Jeena2023}. 
All these aspects would deserve to be explored in future studies, illustrating the various manifestations of rapid rotation in the early universe.

\noindent
BL gratefully acknowledges funding from the Deutsche Forschungsgemeinschaft (DFG, German Research Foundation) under Germany's Excellence Strategy EXC 2181/1 - 390900948 (the Heidelberg STRUCTURES Excellence Cluster). YS and GM have received funding from the European Research Council (ERC) under the European Union's Horizon 2020 research and innovation programme (grant agreement No.~833925, project STAREX).


%

\vspace{5mm}







\bibliography{references}
\bibliographystyle{aasjournal}

\appendix

\section{SFR-UV conversion factors for different star-formation timescales}
\label{sec:kuv_tsf}
\begin{figure*}[h]
    \centering
    \includegraphics[width=0.9\textwidth]{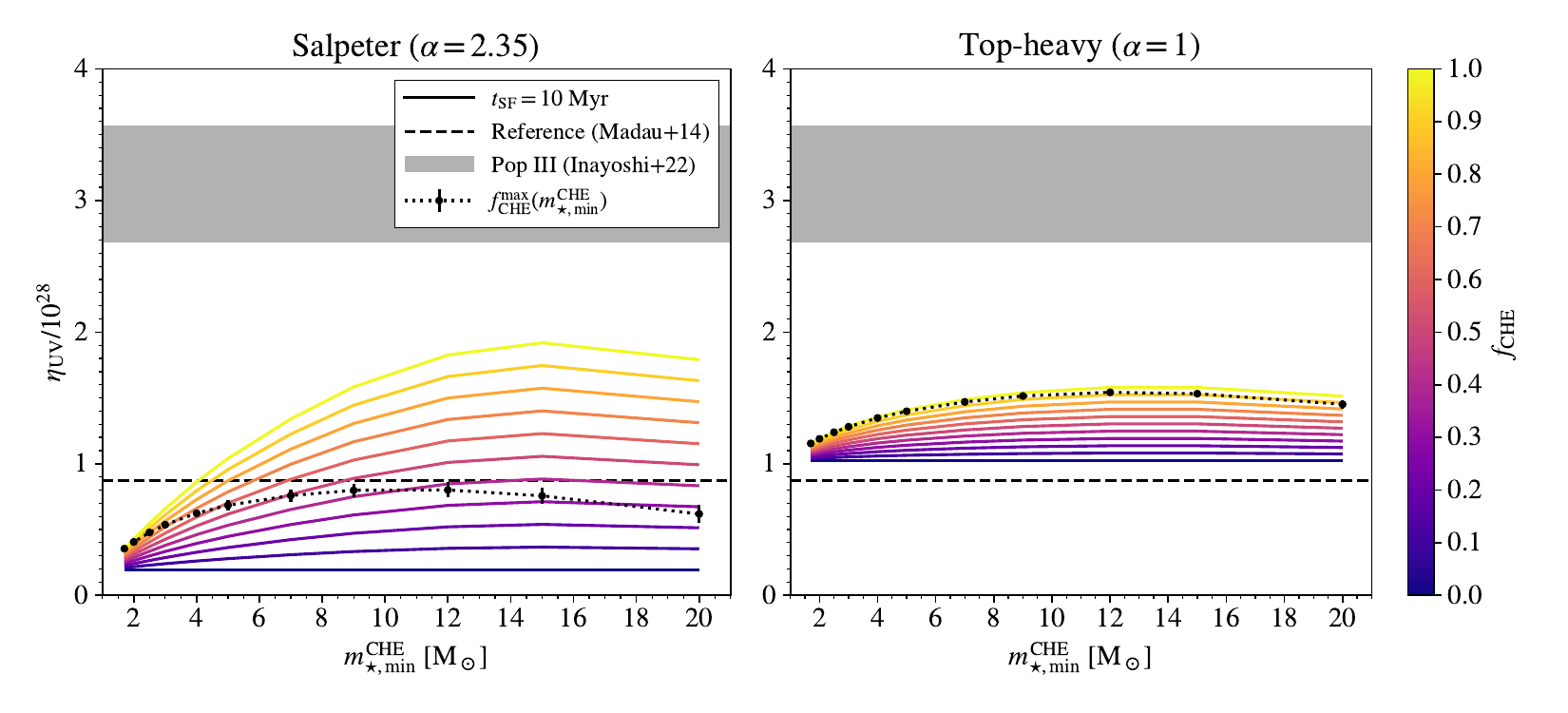}
    \caption{Same as Fig.~\ref{fig:kuv_mmin} but for a star-formation timescale $t_{\rm SF}=10$~Myr.}
    \label{fig:kuv_mmin_tsf10}
\end{figure*}

\begin{figure*}[h]
    \centering
    \includegraphics[width=0.9\textwidth]{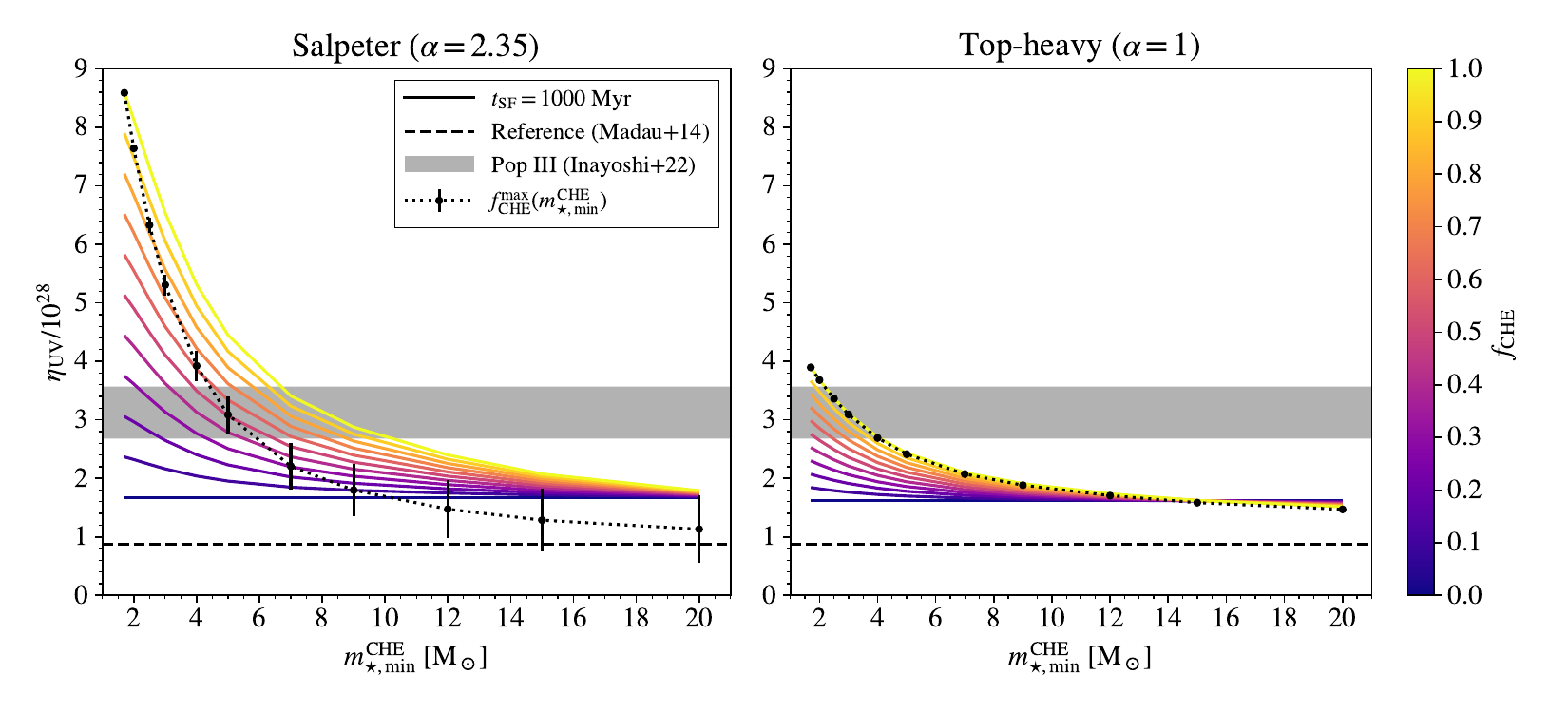}
    \caption{Same as Fig.~\ref{fig:kuv_mmin} but for a star-formation timescale $t_{\rm SF}=1000$~Myr.}
    \label{fig:kuv_mmin_tsf1000}
\end{figure*}

{Figs.~\ref{fig:kuv_mmin_tsf10} and \ref{fig:kuv_mmin_tsf1000} show the results of $\eta_{\rm UV}$ for $t_{\rm SF}=10$ and 1000~Myr, respectively, complementing the results for $t_{\rm SF}=100$~Myr in Fig.~\ref{fig:kuv_mmin}.}

\section{Contribution of nebular emission}
\label{sec:nebular}
\begin{figure*}
\centering
    \includegraphics[width=0.48\textwidth]{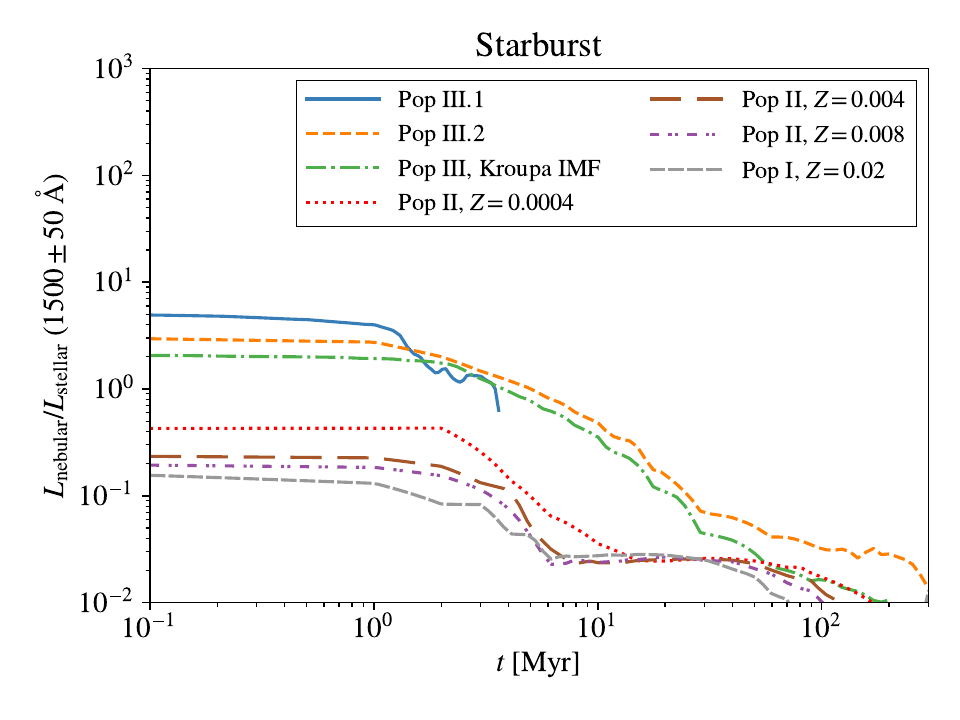}\includegraphics[width=0.48\textwidth]{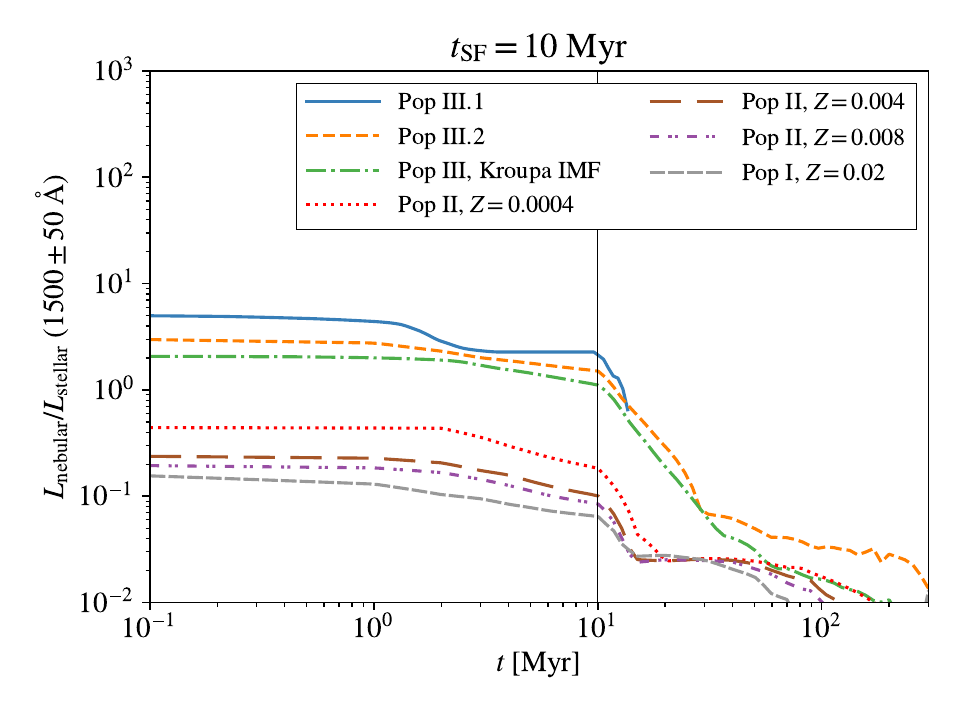}\\
    \includegraphics[width=0.48\textwidth]{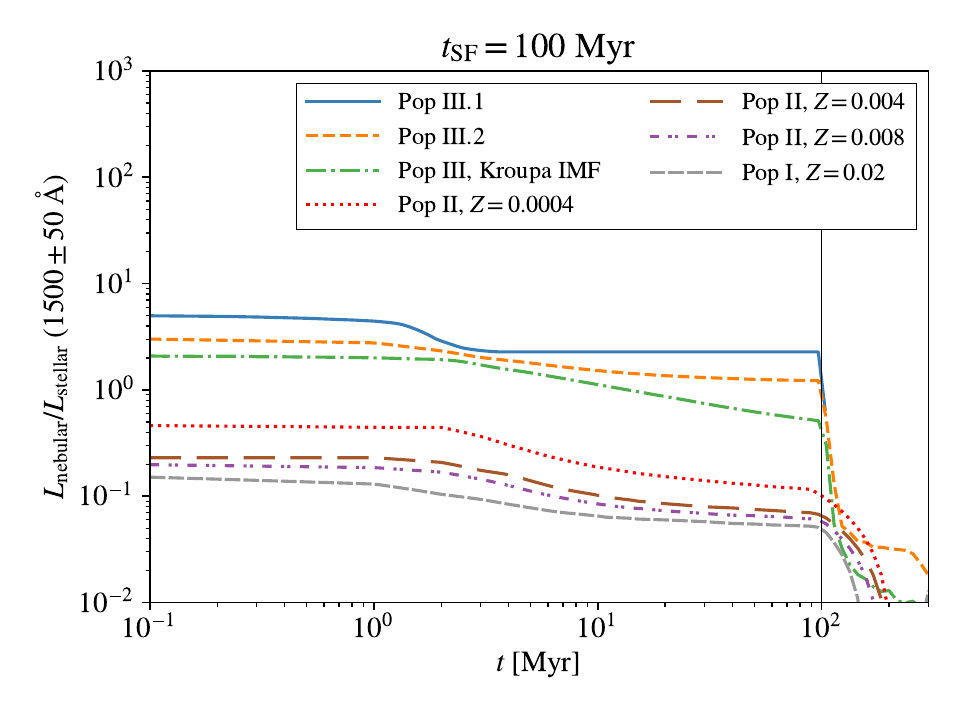}\includegraphics[width=0.48\textwidth]{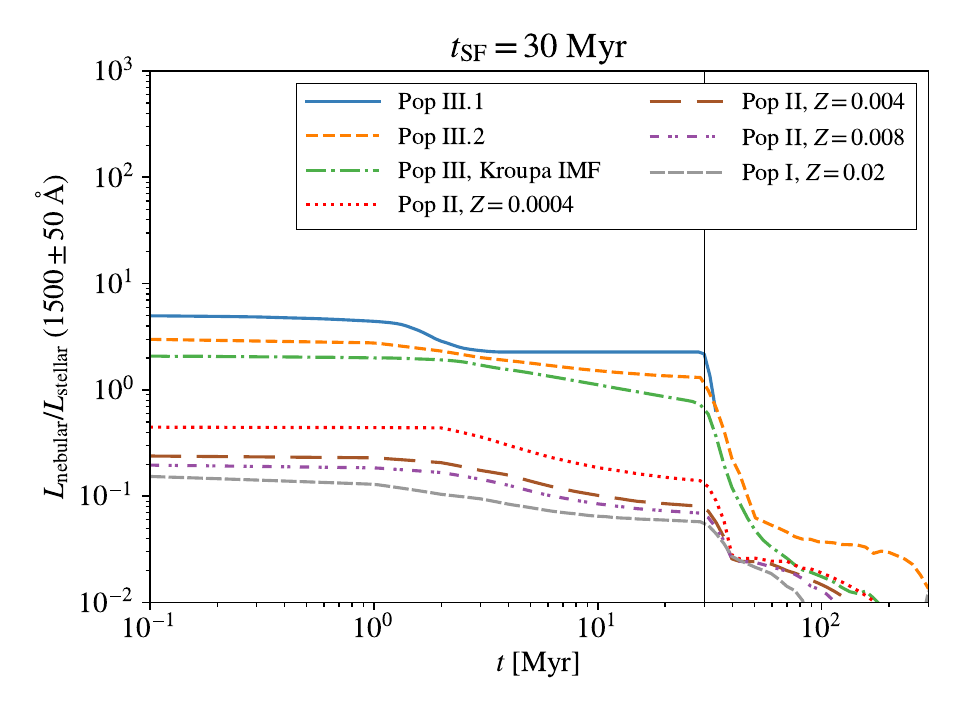}
    \caption{{Ratio between nebular and stellar luminosities in rest-frame UV ($1500\pm 50~\rm\AA$) from an instantaneous starburst and continuous star formation (starting at $t=0$) with a constant rate over timescales $t_{\rm SF}=10$, 30, and 100~Myr (clockwise) for the Yggdrasil models \citep{Zackrisson2011}, assuming no leakage of ionizing photons (optimistic nebular emission). Among these are two models for Pop~III stars with top-heavy IMFs: a Salpeter-like power-law ($dN/dm_\star\propto m_\star^{-2.35}$) in the range $m_\star\in[50,500]~ \rm M_\odot$ (Pop~III.1, solid) and a log-normal distribution with characteristic mass $m_{\star,\rm c}=10~ \rm M_\odot$ and dispersion (in $\ln m_\star$) $\sigma=1$ for $m_\star\in [1,500]\ \rm M_\odot$ (Pop~III.2, dashed), one Pop~III model with a \citet{Kroupa2001} IMF in $m_\star\in [0.1,100]\ \rm M_\odot$ (dash-dotted), three Pop~II models with the same \citet{Kroupa2001} IMF and different metallicities: $Z=0.0004$ (dotted), 0.004 (long-dashed), 0.008 (dash-dot-dotted), and a Population I (Pop~I, densely-dashed) model with the same \citet{Kroupa2001} IMF and $Z=0.02$. For the cases with continuous star formation, we are mainly interested in the results at $t=t_{\rm SF}$ (labeled with vertical lines), which denotes the moment of observation, i.e., when the galaxy has been forming stars for a duration of $t_{\rm SF}$.
    }}
    \label{fig:nebular}
\end{figure*}

{We use the Yggdrasil models \citep{Zackrisson2011} to estimate the possible contribution of nebular emission in rest-frame UV. As shown in Fig.~\ref{fig:nebular}, nebular emission is strongest for very young star-forming clouds, reaching up to $\sim 40\%$ of the stellar luminosity in moderately metal-enriched environments with $Z\ge 0.0004$. However, it decays with time and becomes negligible ($\lesssim 10\%$ of the stellar contribution) shortly ($\sim$ a few Myr $\ll t_{\rm SF}$) after the starburst. Besides, for continuous star formation with a constant SFR over the past $t_{\rm SF}= 100~(10)\rm~Myr$, the nebular luminosity is below $10~(20)\%$ of the stellar luminosity. Note that \citet{Zackrisson2011} only consider non-rotatings stars without CHE and adopt a \citet{Kroupa2001} IMF in the mass range $m_\star\in [0.1,100]~ \rm M_\odot$ for $Z\ge 0.0004$. The nebular emission is stronger at lower metallicities and when the stellar ionizing radiation is enhanced. In the Yggdrasil models of metal-free clouds hosting non-rotating Pop~III stars, nebular emission can dominate the rest-frame UV luminosity in the first few Myr and remain comparable to stellar emission for up to $\sim 20$~Myr after a starburst or continuous star formation with $t_{\rm SF}\sim 10-100$~Myr. Here, the nebular contribution is approximately proportional to the stellar ionizing flux. If this trend holds for moderately metal-enriched ($Z\gtrsim 0.0004$) galaxies most relevant for {\it JWST} observations, nebular emission can become comparable to stellar emission in rest-frame UV under CHE, since the production of ionizing photons is boosted by up to a factor of a few by CHE \citep[Liu et al. in prep.]{Sibony2022}. }


\end{document}